# Dielectric and conductivity relaxation in mixtures of glycerol with LiCl

M. Köhler, P. Lunkenheimer[a], and A. Loidl

*Experimental Physics V, Center for Electronic Correlations and Magnetism, University of Augsburg, 86135 Augsburg, Germany*

**Abstract.** We report a thorough dielectric characterization of the $\alpha$ relaxation of glass forming glycerol with varying additions of LiCl. Nine salt concentrations from 0.1 – 20 mol% are investigated in a frequency range of 20 Hz – 3 GHz and analyzed in the dielectric loss and modulus representation. Information on the dc conductivity, the dielectric relaxation time (from the loss) and the conductivity relaxation time (from the modulus) is provided. Overall, with increasing ion concentration, a transition from reorientationally to translationally dominated behavior is observed and the translational ion dynamics and the dipolar reorientational dynamics become successively coupled. This gives rise to the prospect that by adding ions to dipolar glass formers, dielectric spectroscopy may directly couple to the translational degrees of freedom determining the glass transition, even in frequency regimes where usually strong decoupling is observed.

## 1 Introduction

The glass transition, with its tremendous but continuous slowing down of molecular dynamics when approaching the glass state, still is one of the great mysteries of the physics of condensed matter [1,2]. From an experimental point of view, it is a challenging problem to shed some light on the microscopic processes leading to this phenomenon. During the last decades many experimental investigations have focused on the exceptional dynamics of glassy matter. Recent experimental advances now enable investigating this dynamics in great detail [2]. Especially dielectric spectroscopy plays an important role here as the broad frequency range accessible by this technique allows following the molecular dynamics from the low-viscosity liquid well into the region of the solid glass [3,4]. In addition, there are various processes at frequencies beyond the $\alpha$-relaxation, which found considerable interest in recent years and can be well investigated by dielectric spectroscopy [3,4].

Dielectric spectroscopy is traditionally applied for studying dipolar relaxation phenomena or ionic charge transport. The main contribution to dielectric spectra arises from reorientational motions for the former, and from translational motions for the latter. It is commonly assumed that the freezing of translational motions drives the glass transition (however, also reorientational motions may play an important, so far underestimated role [5,6]). Thus for dipolar glasses, dielectric spectroscopy may suffer from decoupling phenomena. However, in many cases the dipolar relaxation is sufficiently strongly coupled to the structural relaxation determining the glass transition and, thus, its continuous slowing down under cooling mirrors the freezing of the molecular dynamics that leads to the glass state. For example for hydrogen-bonded glass formers as glycerol, this coupling can be rationalized, e.g., by considering that for both, reorientational and translational motions, bonds have to be broken and reformed. In contrast, for the ionic motion in glass formers quite often a complete decoupling from the structural relaxation is found. For example, small ions as Li$^+$ can exhibit significant mobility even in solid glasses. Only in the well-known cases of ionic liquids and melts, i.e. glass formers that are entirely composed of ions (a famous example being $[Ca(NO_3)_2]_{0.4}[KNO_3]_{0.6}$ (CKN)), the ionic motion is usually directly coupled with the glass transition.

The dielectric loss spectra, $\varepsilon''(\nu)$, of glass-formers, made up of dipolar molecules, are dominated by the so-called $\alpha$-relaxation peaks associated with the reorientational motions of the dipoles. Their strong, temperature-dependent frequency shift mirrors the dramatic but continuous slowing down of the molecular dynamics when approaching the glass transition. In the case of glass forming ionic conductors, the dielectric loss spectra are dominated by a divergence towards low frequencies arising from the conductivity contribution of the ions, and information on the structural $\alpha$-relaxation often is not directly accessible. The same can be said for the real part of the permittivity, $\varepsilon'(\nu)$, which is strongly influenced by electrode polarization ("blocking electrodes"). To overcome these problems, Macedo *et al.* [7] proposed the use of the modulus representation for ionic conductors. The complex electric modulus $M^*$ is defined as the inverse of the complex permittivity, $M^* = 1/\varepsilon^*$. Interpreting data in the framework of this representation nowadays is a commonly employed method to obtain information about the charge-carrier dynamics in ionic conductors. In this representation, conductivity and electrode contributions are essentially suppressed. Instead, in the imaginary part $M''(\nu)$ strongly temperature-dependent peaks arise, which can be assumed to be related to the translational ionic motions. The corresponding relaxation time $\tau_\sigma = 1/(2\pi\nu_p)$, with $\nu_p$ the peak frequency, therefore is called conductivity relaxation time.

---

[a] Corresponding author. e-mail: peter.lunkenheimer@physik.uni-augsburg.de

The applicability and correct evaluation of the electric modulus still is controversially debated [8]. Nevertheless, we believe that it is a useful tool for the analysis of data as it effectively leads to a suppression of the phenomena related to dc conductivity and electrode polarization. In addition, the conductivity relaxation times determined from the electric modulus indeed seem to be a good measure of the ionic dynamics and, especially in ionic-melt or -liquid glass formers, can give direct access to the translational dynamics determining the glass transition. For example, for the glass-forming ionic melt CKN, at high temperatures the conductivity relaxation times agree with the structural relaxation times [9]. Only towards lower temperatures, some gradual decoupling of ionic transport and structural relaxation is observed. Similar behavior was recently reported for an ionic liquid [10].

While thus in the $\alpha$-relaxation regime, for both classes of glass formers dielectric spectroscopy often gives rather good access to the translational degrees of freedom, this no longer seems to be the case when considering the dynamics at frequencies beyond the $\alpha$-relaxation. Mainly stimulated by the mode-coupling theory of the glass transition predicting a so-called fast $\beta$-process [11], some works in recent years have focused on the GHz-THz frequency regime. This region was mostly investigated by light and neutron scattering methods. However, for a small number of glass formers also dielectric data were provided [3,12,13]. Among those, the dipolar systems (including glycerol) revealed significant deviations of the dynamic response if compared, e.g., to neutron scattering, which directly couples to density fluctuations, i.e. the translational degrees of motion [3,14,15]. This is valid not only for the regime of the fast $\beta$-process, but may be also the case [16,17] for the Johari-Goldstein $\beta$-relaxation [18] (sometimes called slow $\beta$-relaxation), which is typically detected in the kHz-GHz regime and quite universally observed in all glassy matter. MCT in its basic form treats density fluctuations. Thus, the mentioned discrepancies may arise from the different coupling to translational and reorientational degrees of freedom at high frequencies and the different tensorial properties of the different techniques. Using various extensions of the original MCT formalism, it is possible to understand these differences qualitatively [19], but a quantitative description is difficult.

An alternative approach is trying to achieve identical coupling to density fluctuation in all methods, which according to theory should lead to identical high-frequency response and enable an analysis with basic MCT concepts. For ionic glass formers as CKN this indeed is fulfilled [20]. For dipolar glass formers this purpose may be achieved by adding certain amounts of dissolved ions. While in the pure material dielectric spectroscopy almost exclusively couples to reorientational degrees of freedom, it can be expected that with increasing ion content the ionic dynamics starts to dominate, which for high concentrations should couple to the translational molecular motions. This means that adding ions to a dipolar glass former should enable tuning the coupling of dielectric spectroscopy from reorientational to translational motions. Some support for this course of action arises from the finding that in the ionic-melt glass former CKN the above-mentioned discrepancies in the high-frequency response determined from different methods are not observed [20] as here dielectric spectroscopy directly couples to the ionic motions, i.e. the density fluctuations.

To investigate this notion in more detail, in the present work, we provide dielectric measurements on the prototypical dipolar glass-former glycerol containing varying amounts of LiCl. In the past, various dielectric investigations of solutions of ionic salts in dipolar glass formers were reported (see, e.g., [21,22,23,24,25,26,27]). Many of these works have appeared already rather long ago and are relatively restricted concerning frequency and temperature range and partly treat a limited range of ion concentrations only. Glycerol as solvent is of special interest as it is one of the best investigated dipolar glass formers and structural and dipolar $\alpha$-relaxation appear to be rather well coupled: For pure glycerol, it is well established [28,29] that different methods as, e.g., dynamic specific heat, elastic relaxation, light scattering, as well as dipolar relaxation reveal quite similar $\alpha$-relaxation dynamics [30]. In contrast, at high frequencies, in the region of the fast $\beta$-process, strong deviations from different methods were found [3,14]. Compared to earlier work on LiCl dissolved in glycerol [23], the present investigation covers a significantly broader temperature (204 K < T < 363 K) and frequency (10 Hz < $\nu$ < 1.8 GHz) range and provides information for a larger number of different ion concentration levels. The present work exclusively deals with the $\alpha$-relaxation dynamics; work on the investigation of the fast $\beta$-dynamics currently is in progress and the results will be provided in a forthcoming article, also including information on the Johari-Goldstein process [31] Here we address the question how the ionic dynamics, accessible by the modulus evaluation, and the reorientational (and thus structural) one develop with increasing concentration level. In addition, we provide information on the dependence of the dc conductivity on ion content.

## 2 Experimental details

To record the real and imaginary part of the dielectric permittivity in a broad frequency range, the combination of different techniques is necessary. At low frequencies, 20 Hz < $\nu$ < 1 MHz, stainless-steel parallel-plate capacitors were used and filled with the liquid sample material. The plates were kept at distance using glass-fiber spacers with typical diameters of 100 µm. For the low-frequency range the standard ac-bridge technique using a Hewlett-Packard LCR-meter HP4284 was applied. At frequencies 1 MHz ≤ $\nu$ ≤ 1.8 GHz the impedance analyzer HP4291 was employed using a reflectometric technique, with the sample mounted at the end of a 7 mm coaxial line [32]. For cooling and heating of the samples, a nitrogen gas-heating system was used. For further experimental details the reader is referred to refs. [3], [32], and [33]. The sample materials were purchased from MERCK and measured without further purification. The specified purity for glycerol was ≥ 99.5%. The LiCl concentrations are specified in mol%.

## 3 Results and discussion

We have measured the dielectric loss and dielectric modulus spectra of mixtures of glycerol with ten different concentrations of LiCl ions. The spectra were fitted, employing commonly used empirical functions, namely the Cole-Davidson [34] (CD) function, with an additional dc-conductivity term, for $\varepsilon^*$ and the sum of a CD and Havriliak-Negami (HN) function [35] for $M^*$. All fits were simultaneously performed for the imaginary *and* real part of the dielectric permittivity as well as for the dielectric modulus. The real parts are not shown, as they do not provide significant additional information. The relaxation times resulting from the fits are shown and compared for the different processes. All relaxation times deviate from thermally activated Arrhenius behavior but can be well described with the phenomenological Vogel-Fulcher-Tammann (VFT) law [36]. Ten solutions with LiCl concentrations of $x$ = 0, 0.1, 0.5, 1, 2, 4, 5, 7.5, 10, and 20% were studied. In the following subsection, two representative samples with small (1% LiCl) and high ion concentration (10% LiCl) are discussed in detail.

### 3.1 Dependence of spectra on ion content

Figure 1 shows spectra of $\varepsilon''$ and $M''$ of glass-forming glycerol with 1 mol% LiCl ions, over a temperature range from 204 to 363 K. The dominating feature of the loss spectra is the asymmetrically shaped $\alpha$-peak. In pure glycerol, this peak is ascribed to reorientational motions of the dipoles. As for pure glycerol [3,14,37]. by changing the temperature by about a factor of two, the $\alpha$-peak shifts by more than 10 decades of frequency, which mirrors the tremendous slowing down of the structural dynamics during the transition from the low viscosity liquid to the glass. Obviously, adding 1 mol% of LiCl ions does not change the familiar appearance of the dielectric loss spectra of pure glycerol [3,14,37]. The only difference seems to be observed in the dc-conductivity contribution, which, via the relation $\varepsilon'' \propto \sigma'/\nu$, leads to a steep increase of $\varepsilon''$ towards low frequencies. In Fig. 1(a) this feature is of much higher amplitude than in nominally pure glycerol, where it arises from small amounts of ionic impurities. Figure 1(b) shows the imaginary part of the dielectric modulus for various temperatures. In this representation clearly two peaks can be discerned. It is well known, that a relaxation peak in $\varepsilon''$, as observed in glycerol, leads to a relaxation peak in $M''$, too, however, with a peak position that is significantly shifted to higher frequency [23,24,38]. Thus the peak showing up in $M''$ at higher frequencies is identified with the dipolar $\alpha$-relaxation process. The second, weaker peak, revealed at lower frequencies can be assumed to be related with the translational ion dynamics and mirrors the conductivity relaxation of the mobile ions. The relaxation time corresponding to this process is labeled as conductivity-relaxation time $\tau_\sigma$ [7].

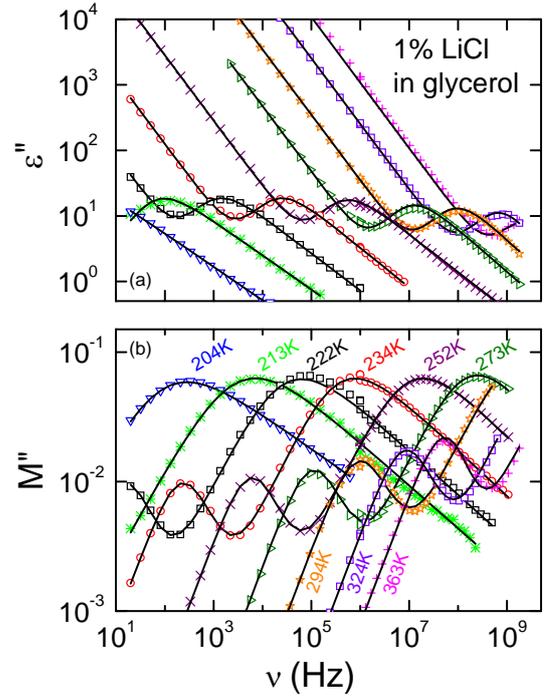

FIG. 1. Frequency dependence of the imaginary part of the dielectric permittivity $\varepsilon''$ (a) and the dielectric modulus $M''$ (b) in glycerol with 1 mol% LiCl at various temperatures (symbols). The measurement temperatures in (a) are the same as denoted for the corresponding curves in (b). The lines in (a) are fits with the CD function and an additional contribution for the dc-conductivity in $\varepsilon''$. The $M''(\nu)$ curves in (b) were fitted by the sum of a CD and a HN function.

While the spectra for glycerol with small LiCl content closely resemble those for pure glycerol, higher ion content changes the appearance drastically. This is demonstrated in Fig. 2 for an ion content of 10 mol% LiCl. In $\varepsilon''$, the $\alpha$-relaxation peak is strongly superimposed by the conductivity contribution. At the lowest frequencies and highest temperatures, deviations from the $1/\nu$ frequency-dependence, arising from ionic charge transport, show up and a flattening of the curves is observed. This behavior is typical for blocking electrodes, i.e. the formation of a space charge close to the sample surface due to the fact that the ions cannot penetrate the metallic electrodes [39]. For the 324 K curve, at $\nu < 10^3$ Hz the slope of $\varepsilon''(\nu)$ increases again. This effect may be ascribed to the fact that there are two species of ions, Li$^+$ and Cl$^-$. The additional increase of $\varepsilon'$ may be due to an extra contribution of the bigger, more immobile chlorine ions, which due to their slower diffusion should show blocking electrode effects at lower frequencies only. In $M''(\nu)$ (Fig. 2(b)), a well developed peak and a second one, showing up as a shoulder at lower frequencies only, are observed. Again, the one at higher frequencies can be identified with the $\alpha$-relaxation peak due to dipolar reorientation. As mentioned before, the frequency shift of the peak in $M''(\nu)$, compared to that in $\varepsilon''(\nu)$, is a well-known feature. The smaller peak at lower frequencies is ascribed to the conductivity relaxation of the translational motions of the mobile ions. Obviously, in contrast to the sample with low ion content, both peaks in $M''$ are no longer well separated, but have nearly merged.

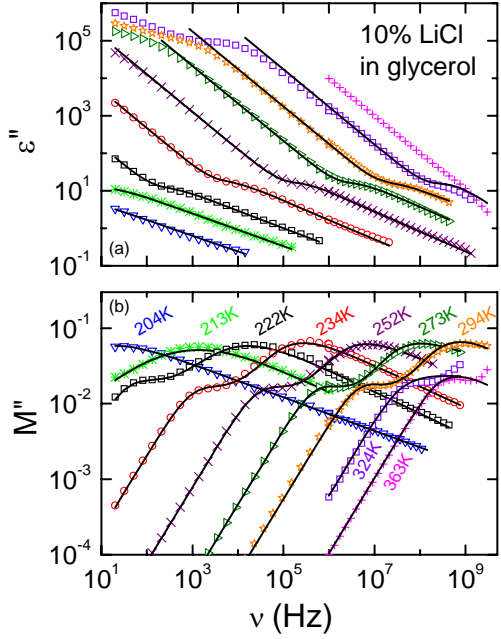

FIG. 2. Frequency dependence of the imaginary part of the dielectric permittivity $\varepsilon''$ (a) and the dielectric modulus $M''$ (b) in glycerol with 10 mol% LiCl at various temperatures. The meanings of the symbols and lines are the same as in Fig. 1.

Figure 3(a) shows the dielectric loss spectra at 234 K for various ion concentrations. In pure glycerol the dipolar $\alpha$-relaxation peak is clearly seen. The increase at lower frequencies is due to conductivity, induced by impurities in the sample. For higher concentrations, a considerable increase in conductivity as well as a significant shift of the $\alpha$-peak to lower frequencies is observed. Only the sample with highest ion concentration (20 % LiCl) deviates from this behavior: No peak or even shoulder can be observed and a clear decrease in the conductivity contribution, compared to the 4 % and 10 % sample, shows up. In Fig. 3(b), the dielectric modulus spectra for the same concentrations and temperature are shown. For the samples with LiCl, the spectra are composed of two maxima, the low-frequency one being due to the translational ion dynamics, whereas the high-frequency one is due to the reorientational motion of the glycerol dipoles. In agreement with the findings of Howell et al. [23], with increasing ion content both peak frequencies successively approach each other and at 20% LiCl content they have almost completely merged. The shifting of the high-frequency peak implies that the ion content strongly influences the reorientational relaxation process, too.

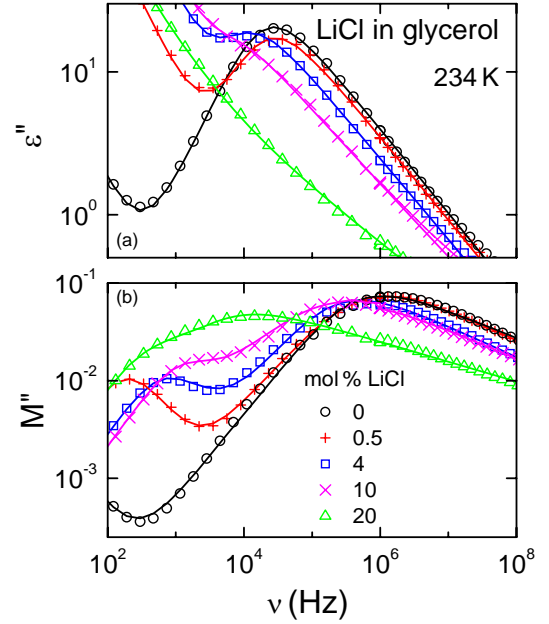

FIG. 3. Frequency dependence of the imaginary part of the dielectric permittivity $\varepsilon''$ (a) and the dielectric modulus $M''$ (b) in glycerol at 234 K for various LiCl concentrations. The lines are fits as in Figs. 1 and 2.

### 3.2 Dc-conductivity, relaxation time and width parameter

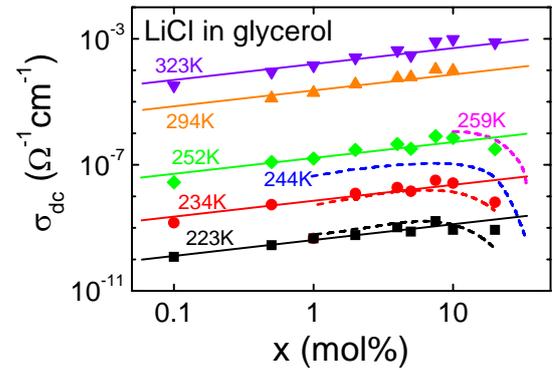

FIG. 4. Dc-conductivity vs. LiCl ion concentration in glycerol for different temperatures. Closed symbols refer to the present results, straight lines illustrate a slope 0.5, dashed lines correspond to literature data from Ref. [23].

Fig. 4 illustrates the dependence of the dc-conductivity $\sigma_{dc}$ on LiCl-concentration at different temperatures. Closed symbols represent the fit values obtained in the present work. The dotted lines denote the results published in Ref. [23], quite well corresponding to our results. In the concentration range from 0.1 to 10% LiCl, the shown solid lines with slope 0.5 describe the experimental data quite well. This implies a square-root behavior of the $\sigma_{dc}(x)$ curves. A weaker than linear increase of $\sigma_{dc}(x)$ is a well known behavior for salt solutions and was also observed for other alkali-halide glycerol mixtures [40]. It can be understood within the theory of Debye, Hückel, and Onsager [41,42,43,44]. As mentioned above, the conductivity of the sample with $x = 20\%$ deviates from the general upward trend of $\sigma_{dc}(x)$, in agreement with earlier findings [23]. Obviously, at high ion concentrations interactions between the ions lead to a strong reduction of ionic mobility.

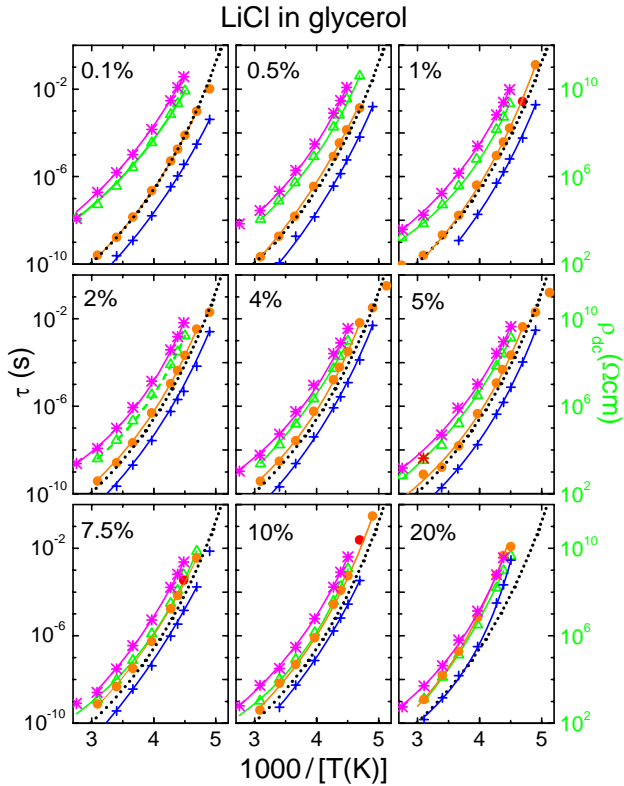

FIG. 5. Temperature-dependent relaxation times $\tau_\varepsilon$ (closed circles), $\tau_\sigma$ (stars), $\tau_R$ (pluses), $\tau_{lit}$ (dotted line) and dc-resistivity $\rho_{dc}$ (open triangles, right ordinate) of all investigated concentrations of LiCl in Arrhenius representation. Lines are fits with a VFT-law. To reduce parameter correlation, $\tau_0$ was held constant for different concentrations. $\tau_{lit}$ corresponds to data on pure glycerol [3].

In the following the relaxation times are labeled like this: $\tau_\varepsilon$ is used for the dipolar $\alpha$-relaxation deduced from the dielectric permittivity spectra. $\tau_\sigma$ denotes the conductivity relaxation time and $\tau_R$ the dipolar relaxation time, determined from the low- and high-frequency modulus peak, respectively. Fig. 5 gives an overview of the complete set of relaxation times determined in this work for all investigated ion concentrations. Solid lines are fits using the VFT equation, $\tau = \tau_0 \exp[DT_{VF}/(T-T_{VF})]$, which provides a good description of the experimental data. $T_{VF}$ denotes the Vogel-Fulcher temperature. The strength parameter $D$ is used in the classification scheme for glass formers, introduced by Angell, to distinguish between strong and fragile glass formers [45]. In Fig. 5, for comparison the dashed lines show the dipolar relaxation times of pure glycerol ($T_{VF} = 129$ K), taken from Ref. [3]. In addition, the dc resistivity $\rho_{dc}$ is provided (pluses, right scale). It is the inverse of the dc conductivity $\sigma_{dc}$, obtained from the fits of $\varepsilon^*(\nu)$ and was scaled to achieve the same number of decades per cm as for the $\tau(T)$ plots (left scale). This quantity was fitted with $\rho_{dc} = \rho_0 \exp[DT_{VF}/(T-T_{VF})]$. Already at first glance, it becomes obvious that all curves shown in the different frames of Fig. 5 approach each other with increasing ion content, with the smallest deviations for 20% LiCl. There is a close relation of $\tau_\sigma$ and $\rho_{dc}$, which both shift nearly parallel to lower values with increasing ion concentration. Such a behavior is expected within the framework of the modulus formalism, where both quantities should be proportional to each other [7]. In Fig. 5, also $\tau_\varepsilon$ and $\tau_R$ shift nearly parallel with varying ion content. Both quantities characterize the dipolar relaxation and it is well known that their ratio is fixed [23,24,38]. Thus the smaller variation of the different curves observed in Fig. 5 for higher LiCl concentrations, is mainly due to the mutual approach of $\tau_\sigma$ and $\tau_\varepsilon$, which are shown in more detail in Figs. 6(a) and (b), respectively.

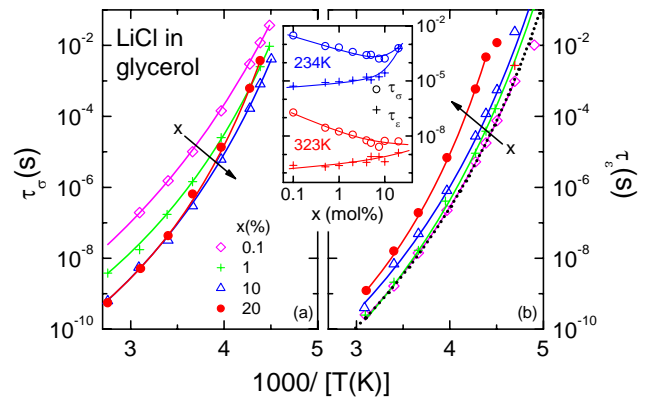

FIG. 6. Conductivity relaxation times $\tau_\sigma$ (a) and dipolar $\alpha$-relaxation times $\tau_\varepsilon$ (b) in Arrhenius representation for selected LiCl concentrations in glycerol. The solid lines are fits using the VFT law. The dotted line in (b) corresponds to literature data for pure glycerol [3]. The inset shows the dependence of both relaxation times on LiCl concentration for two selected temperatures. The lines are guides to the eyes.

Fig. 6(b) shows $\tau_\varepsilon(1/T)$ for selected LiCl concentrations. For low salt concentrations, $\tau_\varepsilon$ quite well corresponds to the relaxation time for pure glycerol (dotted line). Up to $x = 10\%$, a nearly parallel upward displacement of this curve with increasing ion content is observed (see also the pluses in the inset). Similar behavior was also noticed for NaCl-glycerol solutions [27]. It may be ascribed to the increasingly important role of interactions between glycerol molecules and ions, which reduce reorientational mobility [23]. It should be noted that in aqueous solutions of various salts, at relatively low ion concentrations the opposite behavior is observed, i.e. an acceleration of relaxation for increasing ion content [46,47,48]. With further increasing concentrations, however, the relaxation was found to slow down again [46]. At 20 % LiCl content, the $\tau_\varepsilon(1/T)$-curve (Fig. 6(b)) increases significantly stronger than for the lower concentrations, in agreement with the findings of Howell et al. [23]. Within Angell's strong/fragile classification scheme for glass formers [45], this could correspond to a higher fragility. One may speculate that the increased disorder due to the enhanced salt concentration in this sample may lead to a larger density of energy minima in configuration space, which was proposed to be characteristic for fragile glass formers [49]. Alternatively, $T_{VF}$ could be higher than for the other concentrations, which indeed is the outcome of the performed VFT fits (lines in Figs. 5 and 6). However, as in the fits both parameters are highly correlated, no clear statement can be made.

As mentioned before, it may be assumed that the conductivity relaxation times $\tau_\sigma$, obtained from the modulus peaks, provide a characteristic time measure of the ionic motions. In Fig. 6(a), a decrease of $\tau_\sigma$ with increasing LiCl concentration is revealed. The inset of Fig. 6 shows the dependence of $\tau_\sigma$ on ion content for two temperatures (circles). For high ion content, $x > 5\%$, a saturation is observed. Finally, for $x = 20\%$ at low temperatures this behavior seems to reverse and the relaxation times increase again. This overall behavior is in agreement with the findings of Howell et al. [23]. There is no simple argument, why the ionic dynamics should become faster with increasing ion density. For the purpose of the present work, it is sufficient to note that $\tau_\sigma$ approaches the values of the dipolar relaxation time $\tau_\varepsilon$ for high ion concentrations. The $\tau_\sigma(1/T)$ curve for the 20%-LiCl sample seems to exhibit a stronger curvature if compared to the other high ion concentrations. This again may signify a higher fragility of the 20% sample. Also an increase of cooperativity in the ionic motions, due to a stronger coupling at high concentrations may play a role here.

Overall, with increasing ion concentration, $\tau_\sigma$ tends to smaller relaxation times (Fig. 6(a)) whereas $\tau_\varepsilon$ shifts to higher times (Fig. 6(b)) and thus both quantities converge as revealed by the inset of Fig. 6. Finally both $\tau(1/T)$ curves nearly overlap for 20% LiCl as seen in the rightmost lower frame of Fig. 5 (stars and closed circles). While at low ion concentrations, ionic and dipolar dynamics are completely decoupled, this convergence of the timescales of translational ion- and reorientational dipole-motion implies a strong coupling of both dynamics for high ion contents. For pure glycerol, from the agreement of the dipolar $\alpha$-relaxation times $\tau_\varepsilon$ obtained by dielectric spectroscopy with results obtained by other experimental methods [28,29], one can deduce that the reorientational molecular dynamics directly couples to the translational dynamics that determines the glass transition. It may be assumed that this notion also is valid for glycerol with LiCl. This is corroborated by the fact that the glass temperature $T_{g,\varepsilon} = 207$ K for $x = 20\%$, determined from the $\tau_\varepsilon(1/T)$ curve using the condition $\tau(T_g) = 100$ s, agrees reasonably well with the published $T_{g,DSC} = 208.6$ obtained from DSC measurements [23]. Thus, from the present results it can be concluded that for high ion concentrations the ionic motion becomes increasingly coupled to the structural relaxation dynamics, too.

Finally, Fig. 7 provides the width parameter $\beta_{CD}$ as obtained from the CD fits of the dielectric loss peaks. It is well known that in pure glycerol $\beta_{CD}$ increases with temperature as shown by the circles in Fig. 7, saturating at a value below unity [3,4]. At low temperatures, it may well approach a value of 0.5, consistent with the proposed universal exponent 0.5 of the high-frequency flank of the $\alpha$-peak [50]. As becomes obvious from Fig. 7, also for glycerol with LiCl the width parameter increases with temperature. However, for temperatures approaching $T_g$ it reaches much lower values than for the pure material. In addition, a significant decrease of $\beta_{CD}$ with increasing ion content is revealed. Broadening of loss peaks in glass forming matter commonly is ascribed to a disorder-induced distribution of relaxation times [51]. In this context, our results seem reasonable as the ions should introduce additional disorder and thus the distribution should be broadened. Indications for such behavior were also found in earlier works [21,23].

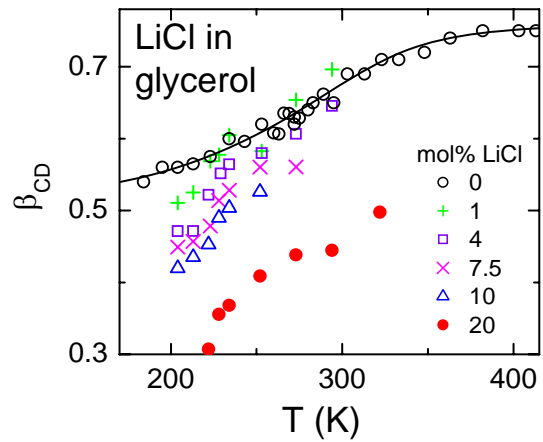

FIG. 7. Temperature dependent width parameter from the CD fits of the loss spectra for selected ion concentrations. The values for pure glycerol were taken from Ref. [4]; the line is drawn to guide the eye. Points are only shown where sufficiently significant information can be provided; especially at high temperatures the superposition by the conductivity contribution hinders an unequivocal determination of $\beta_{CD}$.

## 4 Summary and conclusions

In summary, we have employed dielectric spectroscopy to investigate glass-forming glycerol with various additions of LiCl. The experimental data were analyzed both in the dielectric loss and the modulus representation. With increasing ion content, from a dielectric point of view glycerol turns from a reorientationally to a translationally dominated system. Correspondingly, the single relaxation peak observed in the loss vanishes, being superimposed by a strong conductivity contribution, and a two-peak structure arises in the modulus, merging into a single peak at high concentrations. The ionic dc conductivity increases with a square-root law, which only breaks down for $x > 10\%$. The widths of all peaks increase for higher ion concentrations, due to the stronger disorder introduced by the ions.

We have shown that, while there is a strong decoupling at low salt contents, for high concentrations the translational ion dynamics and the reorientational motions of the dipolar glycerol molecules become directly coupled. In glycerol the latter are itself closely coupled to the structural dynamics, which can be assumed to be dominated by the motion of the centers of gravity of the glycerol molecules, determining, e.g., the viscosity of the system. Thus one can conclude that for high ion contents, the dielectric measurement of the ion dynamics gives direct access to the structural dynamics.

For pure glycerol, in contrast to the $\alpha$-relaxation regime, in the fast $\beta$ regime the reorientational motion of the molecules is known to decouple from the translational one, measured, e.g., by neutron scattering. However, one may speculate that for glycerol with high salt contents the translational motion of the ions may still be coupled to that of the glycerol molecules even at high frequencies. In contrast to the center-of-gravity motion of the glycerol molecules, this ion dynamics is directly accessible by dielectric spectroscopy. Therefore the problem of the different coupling of different spectroscopic methods to density fluctuations arising in the theoretical analysis of the fast $\beta$ regime may be circumvented and the situation may be of similar simplicity as for ionic glass-formers [13,20]. Currently, an investigation of the glycerol-LiCl system at higher frequencies, both with dielectric spectroscopy and with neutron scattering, is under way and the results will be reported in a forthcoming paper [31].